\documentclass[journal]{IEEEtran}

\ifCLASSINFOpdf
 \usepackage[pdftex]{graphicx}
 \DeclareGraphicsExtensions{.pdf,.jpeg,.png}
\else
  \DeclareGraphicsExtensions{.eps}
\fi
\usepackage{tikz}
\usepackage{standalone}
\usepackage{pgfplots}
\pgfplotsset{compat=newest}
\usepackage{amsmath}
\usepackage{array}
\usepackage{algorithm,algorithmic}
\usepackage{csquotes}
\usepackage{subfigure}
\usepackage{xcolor}
\usepackage{hyperref}
\usepackage{cite}
\usepackage{multirow}

\usepackage{academicons}
\usepackage{xcolor}

\usepackage{scalerel}
\usepackage{tikz}
\usetikzlibrary{svg.path}

\definecolor{orcidlogocol}{HTML}{A6CE39}
\tikzset{
  orcidlogo/.pic={
    \fill[orcidlogocol] svg{M256,128c0,70.7-57.3,128-128,128C57.3,256,0,198.7,0,128C0,57.3,57.3,0,128,0C198.7,0,256,57.3,256,128z};
    \fill[white] svg{M86.3,186.2H70.9V79.1h15.4v48.4V186.2z}
                 svg{M108.9,79.1h41.6c39.6,0,57,28.3,57,53.6c0,27.5-21.5,53.6-56.8,53.6h-41.8V79.1z M124.3,172.4h24.5c34.9,0,42.9-26.5,42.9-39.7c0-21.5-13.7-39.7-43.7-39.7h-23.7V172.4z}
                 svg{M88.7,56.8c0,5.5-4.5,10.1-10.1,10.1c-5.6,0-10.1-4.6-10.1-10.1c0-5.6,4.5-10.1,10.1-10.1C84.2,46.7,88.7,51.3,88.7,56.8z};
  }
}

\newcommand\orcid[1]{\href{https://orcid.org/#1}{\mbox{\scalerel*{
\begin{tikzpicture}[yscale=-1,transform shape]
\pic{orcidlogo};
\end{tikzpicture}
}{|}}}}

\usepackage{filecontents}

\makeatletter
\def\ps@IEEEtitlepagestyle{%
  \def\@oddfoot{\mycopyrightnotice}%
}
\def\mycopyrightnotice{%
  \begin{minipage}{\textwidth}
  \centering \scriptsize
  \copyright 2022 IEEE. Personal use of this material is permitted. Permission from IEEE must be obtained for all other uses, in any current or future media, including reprinting/republishing this material for advertising or promotional purposes, creating new collective works, for resale or redistribution to servers or lists, or reuse of any copyrighted component of this work in other works.
  \end{minipage}
}
\makeatother
\begin{document}

\title{Cell-free mMIMO Support in the O-RAN Architecture: A PHY Layer Perspective for 5G and Beyond Networks}

\author{Vida Ranjbar \orcid{0000-0002-7493-3247}\and Adam Girycki \orcid{0000-0003-3683-5790}\and Md Arifur Rahman \orcid{0000-0003-3861-9290}\and Sofie Pollin \orcid{0000-0002-1470-2076}\and Marc Moonen \orcid{0000-0003-4461-0073}\and Evgenii Vinogradov\orcid{0000-0002-4156-0317}
\thanks{Vida Ranjbar, Sofie Pollin, Marc Moonen, and Evgenii Vinogradov, are with Department of Electrical Engineering, KU Leuven, Belgium.}
\thanks{Adam Girycki and Md Arifur Rahman are with IS-Wireless, Pulawska Plaza, ul. Pulawska 45b, 05-500 Piaseczno, Poland.}
\thanks{This research has received funding from the European Union's Horizon 2020 research and innovation programme under Grant Agreement No. 101017171 (MARSAL project).}
\thanks{(Vida Ranjbar and Adam Girycki are co-first authors; Corresponding author: Vida Ranjbar vida.ranjbar@kuleuven.be)}}

\maketitle

\begin{abstract}
To keep supporting next-generation requirements, the radio access infrastructure will increasingly densify. Cell-free (CF) network architectures are emerging, combining dense deployments with extreme flexibility in allocating resources to users. In parallel, the Open Radio Access Networks (O-RAN) paradigm is transforming RAN towards an open, intelligent, virtualized, and fully interoperable architecture. This paradigm brings the needed flexibility and intelligent control opportunities for CF networking. In this paper, we document the current O-RAN terminology and contrast it with some common CF processing approaches. We then discuss the main O-RAN innovations and research challenges that remain to be solved. 

\end{abstract}

\begin{IEEEkeywords}
Cell-free networks, O-RAN, 5G, 6G.
\end{IEEEkeywords}

\IEEEpeerreviewmaketitle
\section{Introduction}
\IEEEPARstart{C}ommunication networks are the nervous system of modern society and the economy. Our world becomes more and more data-centric: the number of sensors is skyrocketing. On the other hand,  
productivity growth relies on creating customer-centric services and automation of manufacturing processes \cite{WhatS6}. Society looks forward to exciting new experiences such as augmented/virtual reality or holographic calls/telepresence \cite{WhatS6,9040264}. 5G made significant progress toward accommodating all these needs and expectations. However, it is predicted that by 2030 this technology will not be able to satisfy the data rate requirements imposed by the fully digitalized world \cite{WhatS6}. The above discussion has motivated academia and industry to start several advanced research activities in order to shape the vision of future 6G networks that are well summarized in \cite{WhatS6,9040264}. 

In this article, among all technological innovations pertaining to 6G, we would like to highlight novel network architectures such as cell-free (CF, or cell-less) massive multiple-input-multiple-output (mMIMO) since this architecture offers high Spectral Efficiency (SE) \cite{7827017} required in 6G \cite{WhatS6}. Moreover, the authors of \cite {8768014} claimed that CF mMIMO networks offer the following benefits: i) large energy efficiency (i.e., reducing carbon footprint); ii) flexible and cost-efficient deployment; iii) uniform Quality of Service (due to absence of cell-edge users and inter-cell interference); iv) favorable propagation conditions (including lower path loss and blockage probability).

There are some fascinating attempts to analyze or even implement CF mMIMO networks using centralized processing (e.g., in \cite{9212395} and \cite{9183752}, respectively). As a result of these works, it becomes clear that this novel communication architecture will need to rely on distributed processing to become scalable and, consequently, feasible \cite{bjornson2020scalable}. This article goes even further in the practical domain. It presents several distributed CF mMIMO architectures aligned with the Open Radio Access Network (O-RAN) vision \cite{ORAN, ORAN_survey}.

The Open RAN movement is primarily led by two cooperating industrial groups: i) Telecom Infra Project (TIP) focused on deployment and execution and ii) the O-RAN ALLIANCE focused on developing and driving standards to ensure that equipment from multiple vendors inter-operate with each other. The O-RAN ALLIANCE  suggests splitting the RAN into smaller components beyond the radio and the baseband unit standardized in 5G. Moreover, O-RAN will open the internal RAN interfaces and allow us to create an open ecosystem where  different vendors will provide different solutions and generate a complete cost-efficient end-to-end ecosystem. This approach is exceptionally operator-friendly due to so-called disaggregation allowing to pick and choose different RAN components from different vendors. Potentially, it would bring more vendors into the wireless industry by allowing smaller companies to occupy their niche in the market. Moreover, the specialization would accelerate technological innovation by upgrading smaller RAN components without waiting for the entire radio or baseband unit to be upgraded.

The main goal of this paper is to introduce how the CF mMIMO architecture links to the O-RAN-specified architecture. We provide a brief overview of CF mMIMO literature and identify implications of the design choices on the O-RAN architecture. Next, the design trade-offs are discussed through a measurement-based performance study using a 64-antenna CF mMIMO deployment.

\section{Cell-free networks}\label{Sec:CF}
CF mMIMO network refers to a network with many distributed Access Points (APs) cooperatively serving User Equipment units (UEs) through coherent joint transmission and reception \cite{7827017} using the same time-frequency resources. Consequently, the concept of cells is eliminated, motivating the name. The APs are connected via fronthaul links to central processing units (CPUs) responsible for the coordination. The CPUs are interconnected by backhaul links. Precoding/combining operations can be performed locally at each AP, centrally at the CPU, or the processing can be distributed between APs and CPUs. 

\subsection{Distributed Processing}
Initial works (e.g., \cite{7827017} by Ngo et al.) assumed a single CPU operating in a Network MIMO fashion (i.e., a centralized approach in which the CPU is responsible for coordinating and processing the signals of all UEs). In this case, APs are quite simple since they only need to receive the signal and transfer it to the central entity (CPU). This approach demonstrated very high SE in simulations \cite{7827017}. Wang et al.~\cite{9183752} put impressive efforts into implementing a cloud-based CF mMIMO network working in a centralized manner. In their computationally advanced testbed, 32 general-purpose servers (with four 18-core processors each) constructed a CPU that can serve up to 16 eight-antenna UEs with 16 eight-antenna APs. 
  
However, the implementation feasibility is questionable for wider-scale networks with more APs and UEs. Indeed, the centralized approach implies that the fronthaul capacity and computational resources, required for each AP to process and share the data signals related to all UEs, grow linearly (or faster) with the number of UEs \cite{bjornson2020scalable}. In other words, the original form of Cell-Free Massive MIMO was unscalable. 

Bj{\"o}rnsson and Sanguinetti~\cite{bjornson2020scalable} suggested a decentralized network architecture in which each (more advanced) AP locally estimates the channels of its associated UEs and uses this information to locally process data signals. In this way, achievable SE is lower but expensive fronthaul, and computational resources are utilized more rationally. In our work, we adopt this idea. However, we also assess the performance of several architectures using centralized processing.

\subsection{Inter-CPU Coordination and Cluster Formation}
Another popular but clearly unrealistic assumption (also made in \cite{7827017,8768014,9183752,bjornson2020scalable}) is utilization of a single CPU. A large-scale CF mMIMO network consisting of APs connected to several CPUs is investigated in \cite{8761828}. 

Inter-CPU coordination is tightly linked to the idea of AP clustering described in \cite{bjornson2020scalable}. In this approach, each UE is served by a subset of APs called "cluster" instead of the full set of APs in the network.  By this, the scalability is improved by confining the signal co-processing within the cluster. Two distinctive clustering approaches may be identified: i) network-centric and ii) user-centric.

\begin{figure}
    \centering
    \includegraphics[width=\columnwidth,trim=270 50 270 70, clip]{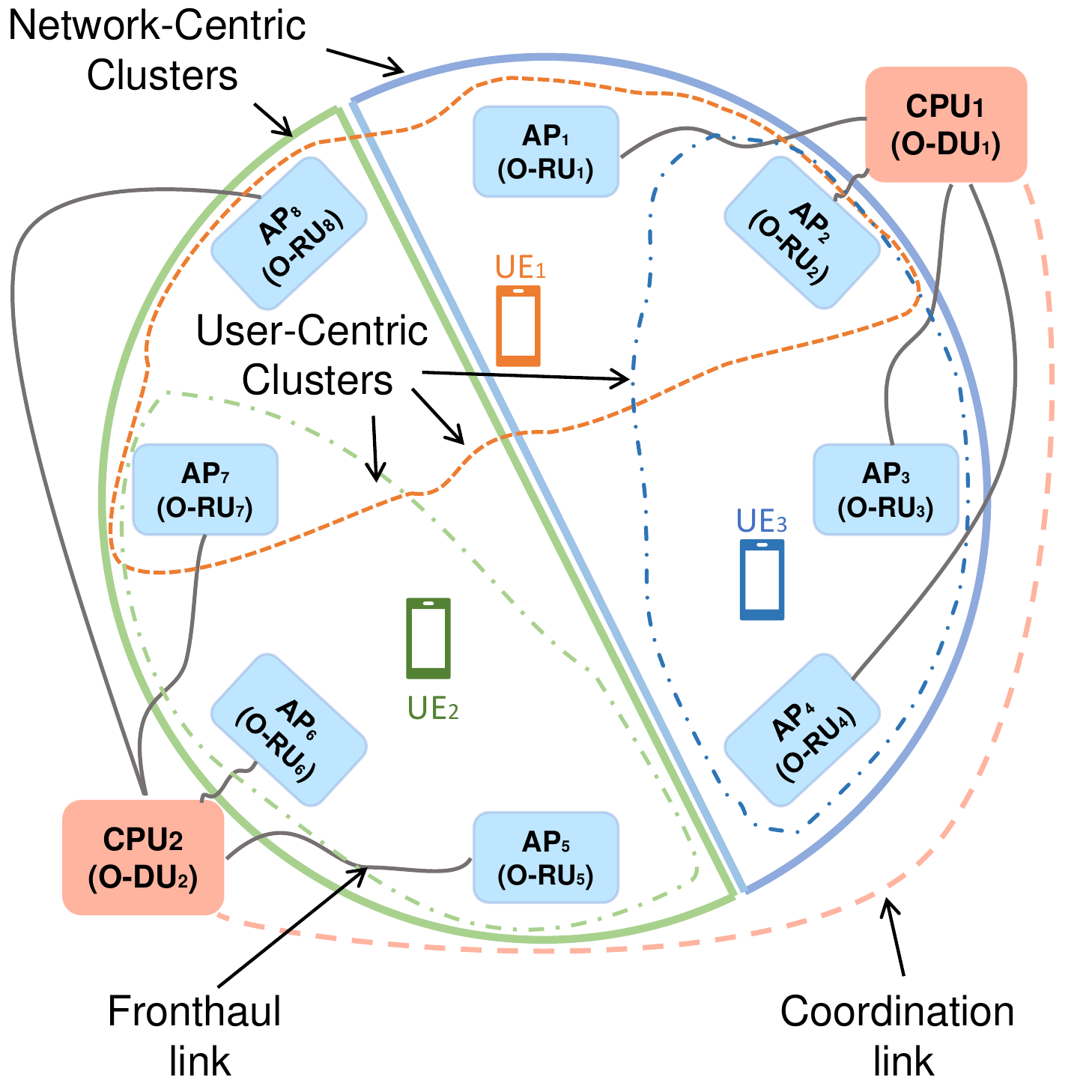}
    \caption{Cell-free system with user- and network-centric clustering and multiple interconnected CPUs. Two types of users are depicted: \textit{local users} (UE$_{2,3}$) and \textit{edge users} (UE${_1}$) served by a single  or multiple network-centric clusters, respectively. The alternative naming of the network nodes (in brackets) is aligned with O-RAN specifications.}
    \label{fig:CF_intro}
\end{figure}

\textbf{Network-centric clustering} (NCC) consists of deploying fixed disjoint clusters of APs serving only the UEs residing in their coverage area (see Fig.~\ref{fig:CF_intro}). Each network-centric cluster is connected to a single CPU. In this configuration, the clusters mutually interfere unless they cooperate (e.g., through coherent transmission which negatively affects the network scalability). Non-cooperative NCC is fully scalable, but it suffers from lower performance as we demonstrate later.

In \textbf{user-centric clustering} (UCC), dynamic clusters of APs are formed based on the needs of each UE. Usually, the user is served by nearby APs offering the best channel (e.g., lowest path loss). Clusters formed in this way may i) partly overlap and ii) consist of APs connected to several CPUs.  UCC results in lower interference; thus it offers better SE than the NCC. However, this concept requires more control signaling to dynamically coordinate several clusters. Moreover, when the number of users (and, consequently, clusters) grows, we may face the need for very intensive inter-CPU coordination when many UEs are served by APs not connected by fronthaul to the same CPU.

Interdonato et al. \cite{8761828} presented a hybrid approach where the CF mMIMO network was shaped by several network-centric clusters. However, each user also forms a cluster in a user-centric manner (APs in such a cluster can belong to different network-centric clusters).

Specifically, for a given UE
\begin{enumerate}
    \item A set of serving APs is identified in a user-centric fashion: they are selected according to some metric (e.g., distance or channel quality).
    \item The clusters involved by the selected APs define the serving network-centric clusters. 
    \item The data to the UE is distributed only to the involved CPUs.
\end{enumerate}

In our work, similarly to the approach in \cite{Zhang&RunhuaNetworkedMIMO}, we consider two types of users: local (served by a single network-centric cluster) and edge (served by two network-centric clusters).

\subsection{Key Design Choices in CF mMIMO Network Design}

Summarizing our reasoning above, we can emphasize the importance of two key points:
\begin{itemize}
    \item Which network unit performs the multi-antenna processing (AP vs.\ CPU)? Which antennas will be used for this processing?  
    \item  Do we consider multiple CPUs and what level of Inter-CPU coordination is acceptable?
\end{itemize}
Answers to these questions define fronthaul and backhaul requirements, computational resources available at the network elements (i.e., APs and CPUs), and the achievable network performance. 

{\color{black}Note that in CF literature, terms precoding and combining are used to describe the multi-antenna processing \cite{bjornson2020scalable}. However, as we demonstrate later in the paper, O-RAN defines functional blocks having the same names but performing different operations. To avoid possible confusion, further we will use the term multi-antenna processing (MAP) to describe the application of precoding and combining vectors to the signals sent in downlink (DL) and received in uplink (UL), respectively. Note that the vectors can be derived by network nodes different from those applying the vectors to the sent/received signals.}  

\section{Cell-Free O-RAN Support}

As it follows from the above reasoning, the CF mMIMO architectures require fine granularity of the processing options (processing at AP vs CPU, Inter-CPU coordination), which can be offered by hardware and software supporting the Open RAN disaggregation concept.
In order to put the CF mMIMO idea out of the academic bubble and make this concept clearer for industry actors, we aim to align the CF mMIMO terminology with the O-RAN alliance vision on communication networks. 

{\color{black}
The "Open RAN" is a movement to disaggregate radio network functions. "O-RAN" refers to the O-RAN Alliance, which publishes new RAN specifications, releases open software for the RAN, and supports its members in the integration and testing of their implementations.   
}

\begin{figure}[t]
    \centering
    \includegraphics[width=0.93\columnwidth]{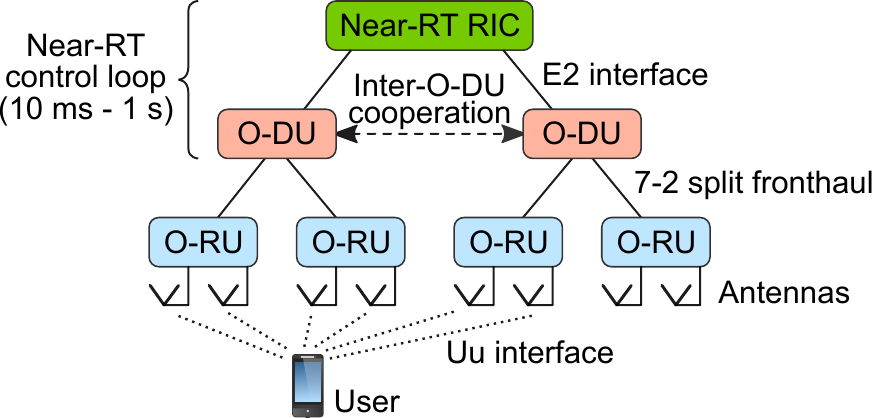}
    \color{black}\caption{The CF mMIMO based on O-RAN architecture. When a UE is served by O-RUs connected to different O-DUs, Inter-O-DU cooperation may be enabled in order to boost the performance.}
    \label{fig:CF_arch}
\end{figure}

{\color{black}
\subsection{O-RAN Architecture and Terminology Alignment}

In O-RAN architecture, the Next Generation Radio Access Network (NG-RAN) \cite{Rel16} is disaggregated into O-RAN Central Unit (O-CU), O-RAN Distributed Unit (O-DU) and O-RAN Radio Unit (O-RU) \cite{ORANWG1}. O-CU is further split into O-CU Control Plane (O-CU-CP) and O-CU User Plane (O-CU-UP). Multiple O-CUs and O-DUs are connected to the Near-Real Time RAN Intelligent Controller (Near-RT RIC) for centralized NG-RAN performance control. 

Consequently, now we can associate the CF mMIMO network components presented earlier with the O-RAN nodes. The CPU is represented by the O-DU while the AP is represented by O-RU. Note that in the following, we will use this terminology instead of the one used in \cite{7827017,8768014,9212395,9183752,bjornson2020scalable,8761828}.

We think that O-RAN architecture will enable practical implementations of future CF mMIMO architectures, as shown in Fig.~\ref{fig:CF_arch}.  
Note that if several O-DUs are involved in serving a given user, then the serving O-DU should be selected. It is responsible for the Medium Access Control (MAC) signaling exchange with the user and holds the upper Radio Link Control (RLC) protocol layer where the user data is buffered.
}

\subsection{O-RAN PHY Layer Split }
The CF mMIMO MAP takes place in the PHY layer. 
The O-RAN splits the PHY layer, placing some PHY functions in the O-DU and some in the O-RU. Therefore, these two nodes and interfaces between them are the focus of our interest.

\begin{figure}[H]
    \centering
    \includegraphics[width=0.49\textwidth]{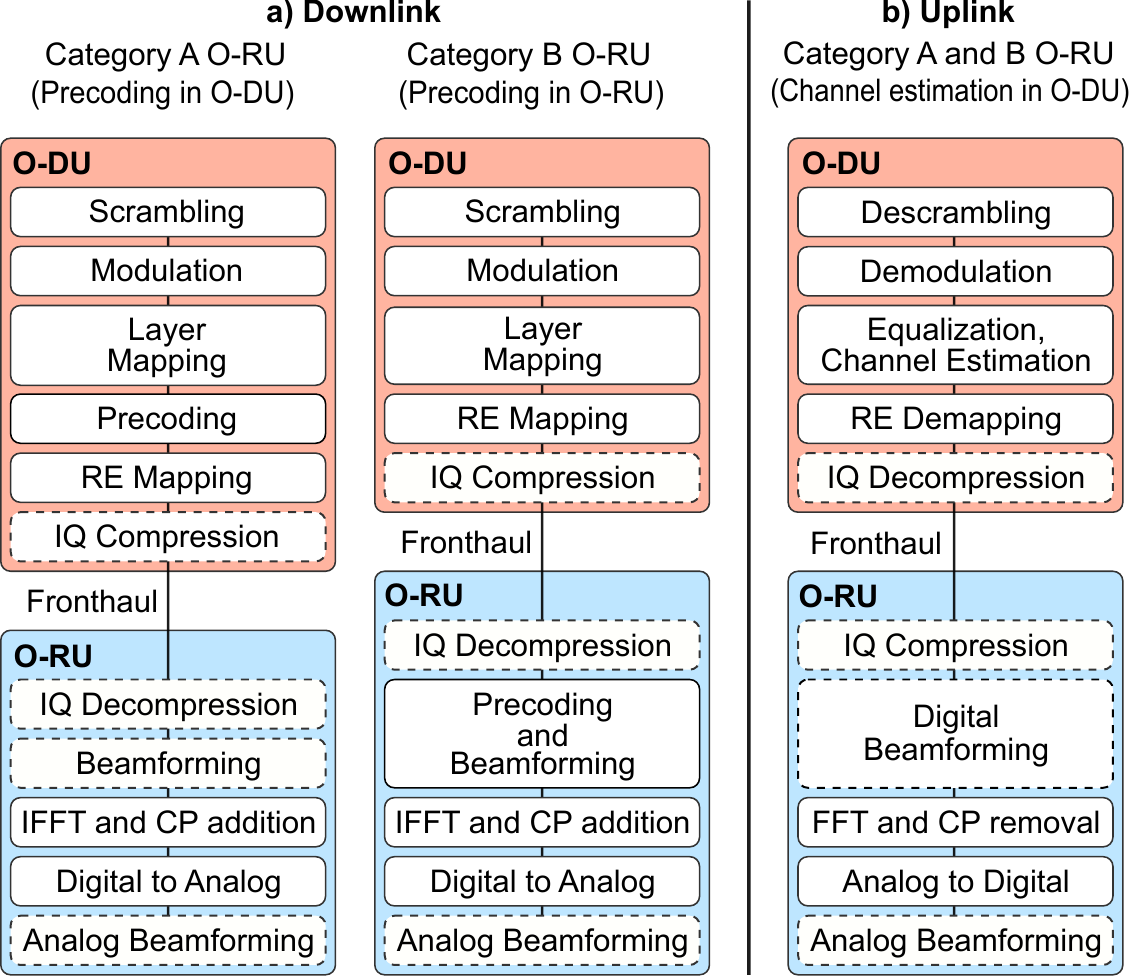}
    \caption{Functionality of the O-RUs. Dashed line indicates an optional block. Abbreviations: Resource Element (RE), Fast Fourier Transform (FFT), Inverse Fast Fourier Transform (IFFT), Cyclic Prefix (CP),  In-phase and Quadrature components (IQ).}
    \label{fig:split_7.2}
\end{figure}

When selecting the PHY layer functional split, there is a trade-off between:
\begin{itemize}
    \item Keeping O-RUs as simple as possible to reduce the O-RU cost and power consumption.
    \item Having the fronthaul interface at a higher level to reduce the interface throughput.
\end{itemize}
To resolve this issue, O-RAN has selected a single split option known as 7-2x, locating (de-)modulation, channel estimation, and equalization in the O-DU, while the digital beamforming is performed by O-RUs.
Additionally, the O-RAN split 7-2x considers two types of O-RUs: advanced O-RUs (Category B), which can execute the precoding operation, and simpler O-RUs (Category A) that do not have this functionality. 
A more detailed description of the operations performed by O-RUs of different categories (and their O-DUs) is shown in Fig.~\ref{fig:split_7.2}.

\subsection{Multi-antenna processing for CF mMIMO in O-RAN}\label{sec:Multi-antenna_processing}
The key function in the CF mMIMO is the MAP, which aims at improving the quality of the received signal at the intended node while reducing interference to non-intended nodes. The MAP is carried out in DL and UL by inverse operations:
\begin{itemize}
\item In DL, for each user, a complex scalar symbol of user data is multiplied by the user's MAP vector of the length equal to the number of O-RU antennas. This operation results in a vector of the same size as the number of antennas (known as precoding vector in CF literature). Then vectors computed for each user are summed up, and the resulting vector represents the complex numbers that the antennas will send in the DL.
\item In UL, in order to detect the signal sent by the user, a vector of symbols received from antennas is multiplied by the MAP vector specific for the user (the operation is also referred to as the combining).
The same process is applied for each user using a user-specific MAP vector. 
\end{itemize}

The 7-2x split fronthaul scales with the number of layers, not antennas. Therefore, we see two basic O-RU configurations possible with respect to the placement of the MAP:
 \begin{enumerate}
    \item {\em O-RU equipped with multiple antennas:} MAP can take place in "Beamforming" (Category A O-RU) or "Precoding and Beamforming" (Category B O-RU Category) functional block for DL (Fig.~\ref{fig:split_7.2}a) the "Digital Beamforming" functional block for UL (Fig.~\ref{fig:split_7.2}b). In this case, we rely on channel-information-based beamforming \cite{ORAN_survey}.
    \item {\em O-RU equipped with a single antenna:} MAP can take place in the  "Precoding" functional block in the O-DU for DL, and the "Equalization, Channel Estimation" functional block in the O-DU for UL.
\end{enumerate}

The per-user MAP vectors (e.g., derived using the Minimum Mean Square Error [MMSE] method) are concatenated into a two-dimensional MAP matrix:

\begin{itemize}
    \item First dimension (rows) is formed by a set of inter-working antennas.
    We have investigated three levels of inter-working antennas:
\begin{itemize}
    \item O-RU antennas.
    \item O-DU antennas (i.e.,\ antennas of O-RUs connected to the O-DU).
    \item All antennas (called global processing).
\end{itemize}
    
\item Second dimension (columns) is created by users served by the set of inter-working antennas:
    \begin{itemize}
        \item Local users. 
        \item Local and edge  users. 
        \item All users in the network.
    \end{itemize}
\end{itemize} 

\subsection{User Categorization for Clustering in O-RAN}
The O-DU hosts the scheduler, which allocates radio resources for UEs. Upon scheduler decision, the user data is sent to the O-RU for transmission, or data is received at the O-RU.
In the CF mMIMO, transmission/reception takes place simultaneously over multiple O-RUs, as shown in Fig.~\ref{fig:CF_arch}. 
Thus, from the O-DU point of view, UEs can be divided into two categories: 
\begin{itemize}
    \item {\em Local users}, which can be satisfyingly served by O-RUs connected to one O-DU.
    \item {\em Edge users}, which should be served collaboratively by O-RUs connected to two (or more) O-DUs.
\end{itemize}
Serving the edge users requires inter-DU cooperation. 
In the next section of the paper, we show that serving edge users with O-RUs of multiple O-DUs improves the user SE.

\subsection{Inter-O-DU coordination interface}\label{sec:O-DU_interworking}
The Inter-O-DU coordination is critical for enabling the full potential of the CF mMIMO networks deployment.
The coordination can be supported through the Inter-O-DU coordination interface (see Fig.~\ref{fig:CF_arch}) which is currently not defined in the O-RAN standard. Specification of the Inter-O-DU cooperation interface signaling is not straightforward and requires further investigation. In this work, we describe some basic concepts of the interface.

The main purpose of the interface is to serve the user by O-RUs connected to more than one O-DU which is especially advantageous for edge users.
One of the cooperating O-DUs should be considered a serving O-DU. The Inter-O-DU cooperation interface can be used to both exchange user data between O-DUs as well as send necessary signaling required in the MAP derivation. The presence of the interface may improve UE SE, which will be shown in the next sections.

\subsection {CF mMIMO Deployment Options}
As mentioned in  Section~\ref{sec:Multi-antenna_processing}, when deriving the MAP vector, two dimensions are considered: a) Size of the interworking antenna set (O-RU, O-DU, or global); b) Absence or presence of the Inter-O-DU coordination interface.

Based on the above aspects we have identified five CF mMIMO deployment options, which are defined in Table~\ref{tab:options}.

\begin{table}[htb]
\color{black}
\caption{CF mMIMO deployment option definition.}
 \label{tab:options}
\centering
\begin{tabular}{|c|c|c||c|}
\hline
Deployment & Inter-working & Inter-O-DU & O-RAN\\ 
option & antennas & coordination & support\\ 
\hline\hline
Option 1 & O-RU & Absent & Yes \\ \hline
Option 2 & O-RU & Present & No\\ \hline
Option 3 & O-DU & Absent & Yes  \\ \hline
Option 4 & O-DU & Present & No  \\ \hline
Option 5 & Global & --- & Yes \\ \hline
\end{tabular}
\end{table}

It is worth mentioning that i) MAP vectors are computed jointly for the inter-working antennas and ii) users can be served by two or more sets of independent inter-working antennas constructing a complete set of serving antennas (e.g., when a user is served by a cluster consisting of several O-RUs calculating their MAP matrices independently from each other).

\begin{itemize}
    \item In options 1 and 2, the MAP vector is computed separately for each O-RU neglecting the channels between the user and antennas of other O-RUs. 
    \item In options 3 and 4, the MAP vector used by each O-RU for a particular user is computed jointly for all O-RUs of one particular O-DU (O-DU level information). This calculation is computationally more demanding but results in MAP vectors providing higher SE as we will demonstrate later in the article. 
\end{itemize}

In this work, we assume that the total number of users is constant. However, the presence or absence of the {\em Inter-O-DU cooperation} interface determines the number of users served by O-RUs:
\begin{itemize}
    \item Without the Inter-O-DU coordination interface: all O-RUs serve only the users located within the same O-DU (i.e., all users are "local").
    \item With the Inter-O-DU coordination interface: i) all O-RUs of the O-DU serve local users; ii) additionally, the O-RUs serve edge users of neighboring O-DUs. Specifically, in the UL, the serving (or "primary") O-DU of an edge user receives the other cooperating O-DUs' estimate of the user's signal through the interface and tries to refine its own estimate of the edge user's signal with that information. In the DL, the serving O-DU sends the intended signal for the edge users to the other cooperating O-DUs so all the O-RUs associated with the cooperating O-DUs process and transmit the user signal. 
    This will improve the UE SE, but at the cost of increased fronthaul and signaling load and complexity of calculations.
\end{itemize}
Since the Inter-O-DU cooperation interface is not supported by the O-RAN, deployment options 2 and 4, which require the presence of Inter-O-DU coordination, cannot be implemented in alignment with the current version of the O-RAN architecture (see Table~\ref{tab:options}). On the other hand, O-RAN can support deployment options 1 and 3, which do not require an Inter-O-DU coordination interface. 
Both Category A and B O-RU can be used for CF mMIMO.

Option 5 may be implemented by connecting all O-RUs to one global O-DU where the global processing is performed. O-RAN can support this, but it is impractical due to high requirements on the O-DU hardware resources (processing, storage, and network).
We can envision another implementation of option 5 with several O-DUs where signals received by all antennas are exchanged between O-DUs through the Inter-O-DU interface. 
However, this is not supported by the O-RAN due to the lack of an Inter-O-DU coordination interface. Moreover, this information exchange would result in a very high Inter-O-DU coordination interface and fronthaul load since all users would be served by all antennas.

\subsection{Near Real-Time RAN Intelligent Controller}
{\color{black}
In the previous section, we have identified two important configuration aspects which may impact the CF mMIMO performance: the size of the inter-working antenna set and the presence of the Inter-O-DU coordination interface. In the case of the presence of the Inter-O-DU interface, edge users will be served by O-RUs of more than one O-DUs. Therefore, users need to be categorized into local and edge users. Next, the MAP vector must be calculated. Thus, the CF mMIMO controlling process can consist of two steps:
\begin{enumerate}
    \item User categorization, which can be based on the Channel State Information (CSI) measurements collected from the O-RUs. 
    The user categorization aims at the determination of the serving O-DU and, when the Inter-O-DU-coordination interface is present, categorizing the user into local or edge users.
    \item Determination of the MAP vector for each O-RU serving the UE. 
\end{enumerate}

The above controlling steps should be repeated frequently enough to follow changing radio environment and system load. 
In O-RAN architecture, the Near-RT RIC controls the O-DUs through the E2 interface (see Fig.~\ref{fig:CF_arch}). Its control loop time, defined by the O-RAN standard \cite{ORANWG1} from 10~ms to 1~s, makes Near-RT RIC a suitable node for the user categorization described in step I. Though Near-RT RIC may also be used for the MAP vector computation, O-DUs are more suitable for step II as they have the real-time control of O-RUs.

The number of O-DUs, which can be connected to one Near-RT RIC, depends on the E2 interface link capacity and the E2 interface signaling load. 
The controlling algorithms located in the Near-RT RIC are referred to as the {\em xApps}. 
A {\em mMIMO CF xApp} will collect the CSI for all users and will use its intelligent algorithms to select the serving O-DU as well as user category.
With its mMIMO CF xApp, which selects the O-RUs serving each user, the Near-RT RIC can support any deployment option defined in the previous section.
The deployment options determine the MAP matrix size and the Near-RT RIC can control any of them.

When the mMIMO CF xApp also selects the MAP vector at step II, the Near-RT RIC does not limit the selection of deployment options either. In the case when the RIC defines MAP vectors, the O-RU-, O-DU-, or network-level MAP can be used depending on the available CSI and hardware resources. However, this does not imply that the presence of the Near-RT RIC directly enables the implementation of option 5. Apart from the global MAP vector calculation (that Near-RT RIC can provide), option 5 also requires the user data to be sent from the serving O-DU to neighboring O-DUs. As the Near-RT RIC is a signaling node, it does not support sending user data. Therefore, in networks with multiple O-DUs, option 5 requires an Inter-O-DU coordination interface for user data exchange or, alternatively, the O-RU to be connected to multiple O-DUs. Both of the solutions are not supported by the current O-RAN architecture.
}

\section{Experimental Performance Estimation of O-RAN Cell-Free Deployment Options}
In this section, we quantify the performance of various deployment options discussed above. We base the analysis on CSI obtained in a dense CF mMIMO deployment \cite{9129126}. The dataset is obtained with the KU Leuven testbed deployed in an indoor environment. 
The central frequency is 2.6~GHz and system bandwidth is 20~MHz.

\begin{figure}[t!]
       \includegraphics[width=0.49\textwidth]{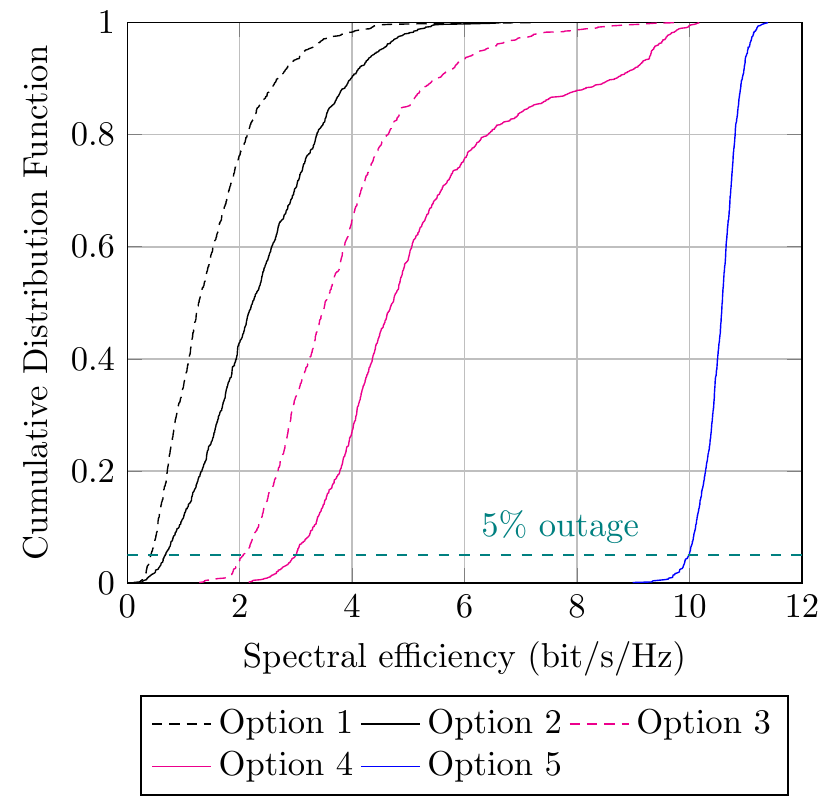}
		\caption{Cumulative Distribution Function of subcarrier SE.}
		\label{fig_CDF}
\end{figure}

Fig.~\ref{fig:CF_intro} represents the scenario: we deployed 8 O-RUs with 8 antennas each (64 antennas in total). We assume two O-DUs each of which serves 4 O-RUs.
The O-RUs connected to the same O-DU compose a cluster, thus we have two network-centric clusters.
This CF mMIMO network is used to serve {\color{black} simultaneously} 10 UEs (both local users and edge users). Each UE occupies the whole 20 MHz bandwidth. We run the simulation for 100 different setups (i.e.,\ different location of the users), and the measured 1000 ($10\times100$) SE realizations are used to produce Fig \ref{fig_CDF}.

To illustrate the performance of different levels of cooperation between O-DUs, we investigate the five deployment options described in the previous section. Note that among multiple possible methods, we use MMSE method to compute {\color{black} MAP vectors} in all the deployment options. 

In Fig. \ref{fig_CDF}, the five {\color{black}deployment options} are compared in terms of their UL SE. 
The horizontal line indicates the worst service among the majority of the users ($95\%$), called $5\%$ outage\cite{nayebiLSFD}.
As observed in the Fig. \ref{fig_CDF}, decreasing the amount of centralization or the number of inter-working antennas in Table \ref{tabII} (i.e., option 5 $\rightarrow$ options 4 and 3$\rightarrow$ options 2 and 1) adversely affects users' service in general. This is because, in options 1 and 2, the MAP vectors are calculated assuming only O-RU local information, but in options 3 and 4, the calculation is based on the O-DU's information of the network, which includes all the information it receives from its O-RUs.
Option 5 is yet a higher level of centralization when compared to options 3 and 4, and as expected, it will further improve the performance, here in terms of SE.
 Moreover, as it is seen in Fig. \ref{fig_CDF}, the advantage of option 2 over option 1 and option 4 over option 3 is due to the O-DU collaboration interface for edge users. 
The comparison summary can be found in Table~\ref{tabII}. In the last column of this table, the $5\%$ outage improvement of different options in comparison with option 1 is shown.

\begin{table}[h]
\color{black}
\centering
\caption{Performance of O-RAN CF mMIMO networks depending on the deployment option.}
\label{tabII}
\begin{tabular}{|c|c|c|c||c|}
\hline
 & & \multicolumn{2}{c||}{Serving} & \\
Deployment & Inter-working & \multicolumn{2}{c||}{antennas} & $5\%$ outage\\ \cline{3-4}
option & antennas & Local & Edge & improvement\\ 
& & user & user & \\ 
\hline\hline
Option 1 & 8  & 32 & 32 & $\times$1\\ \hline
Option 2 & 8 & 32 & 64 & $\times$1.6\\ \hline
Option 3 & 32 & 32 & 32 & $\times$4.9\\ \hline
Option 4 & 32 & 32 & 64  & $\times$7.15\\ \hline
Option 5 & 64 & 64 & 64 & $\times$23.8\\ \hline
\end{tabular}
\\[3mm]
\end{table}

\section{Conclusions}
This paper has mapped the CF mMIMO nodes terminology onto the O-RAN architecture. Moreover, it explained multiple options to implement CF mMIMO in the current or possible future O-RAN generations. More precisely, the key conclusions can be summarized as:
\begin{itemize}
    \item Current O-RAN architecture supports Cell-Free networks not requiring the Inter-O-DU interface.   
    \item Inter-O-DU interface should be specified in O-RAN in order to enable Cell-Free networks offering high Spectral Efficiency.
    \item Near-RT RIC is the most suitable node for defining the set of O-RUs (cluster) serving a given user. 
    \item O-DU is the most suitable node for calculating user-specific Multi-Antenna Processing vectors (also known in Cell-Free literature as precoding/combining vectors). Near-RT RIC also could perform this operation but with a more significant delay (10~ms to 1~s instead of 1~ms).
    \item The Multi-Antenna Processing vectors can be applied both in O-RU and O-DU nodes. However, a multi-antenna O-RU requires applying the vectors in the O-RU exclusively.
\end{itemize}

Future work should focus on extending the current spatial scheduling work to joint time/frequency and spatial domain scheduling strategies. The impact of the allocation of O-RU to UE for different scheduling strategies should be investigated. This paper presented simple procedures to decide on local or edge UE categories. A more thorough investigation is needed to come to solid and efficient solutions for practical cell-free systems. 

Finally, the presented work should be extended to multi-band approaches: the UE--O-RU association might be frequency-dependent. Multiple O-DU coordination approaches, with different levels of coordination for different frequency bands, also require more study. 
The degrees of freedom for future, open, disaggregated RAN are immense. Intelligent and data-driven methods will be needed to tune all parameters and achieve the best possible configuration of the network resources at each time, for each UE and each different application.

\bibliographystyle{IEEEtran}
\bibliography{Refs}
\section*{Biographies}

\textbf{Vida Ranjbar} (vida.ranjbar@kuleuven.be) received her master's degree in electrical engineering from University of Tehran in 2017. She is currently pursuing her Ph.D. studies at the electrical engineering department, ESAT, KU Leuven. Her research interest includes signal processing in cell-free massive MIMO systems.

\textbf{Adam Girycki} (a.girycki@is-wireless.com) holds the position of R\&D Expert in IS-Wireless, Piaseczno, Poland, and is particularly expert in the field of 2G, 3G, 4G, and 5G Radio Network Design, Signaling, and Optimization. Recently, he has also enrolled in a joint Ph.D. degree program with KU Leuven, Belgium. He graduated in Technical Physics (1994) from the Silesian University of Technology in Gliwice, Poland, and Completed his Ph.D. studies in Electronics and Telecommunications (1998). Since 1999 has been a certified technical instructor of Ericsson. In 2002, he established ENKI telecom training company.

\textbf{Md Arifur Rahman} (a.rahman@is-wireless.com) is an R\&D Expert in IS-Wireless, Piaseczno, Poland, focusing on solving the radio resource management (RRM) problems of future wireless networks. Currently, he is the team leader of an H2020 project called MARSAL. He holds a Ph.D. degree in Electrical Engineering from University of Ulsan, South Korea in 2019. Before he joined IS-Wireless, he worked as a Postdoctoral Researcher at CentraleSupelec, Campus of Rennes, France. His research interests include cell-free mMIMO, cloud computing in 5G networks, multi-access edge computing, virtual radio access networks, AI application in wireless communications, digital twins, and emerging technologies in wireless communications.

\textbf{Sofie Pollin} (sofie.pollin@kuleuven.be) received the Ph.D. degree (Hons.) from KU Leuven, in 2006. From 2006 to 2008, she continued her research on wireless communications, energy-efficient networks, cross-layer design, coexistence, and cognitive radio at UC Berkeley. In 2008, she returned to imec to become the Principal Scientist at the Green Radio Team. She is currently a Full Professor with the Electrical Engineering Department, KU Leuven. Her research interests include networked systems that require networks that are ever more dense, heterogeneous, battery-powered, and spectrum-constrained. Sofie is a BAEF Fellow and a Marie Curie Fellow.

\textbf{Marc Moonen} (marc.moonen@kuleuven.be) is currently a Full Professor with the Electrical Engineering Department, KU Leuven (Belgium). He was the Chairman of the IEEE Benelux Signal Processing Chapter from 1998 to 2002, a Member of the IEEE Signal Processing Society Technical Committee on Signal Processing for Communications, the President of the European Association for Signal Processing (EURASIP), and a Fellow of EURASIP and IEEE. Currently, Marc is heading a research team working in the area of numerical algorithms and signal processing for digital communications, wireless communications, DSL, and audio signal processing.

\textbf{Evgenii (Genia) Vinogradov} (evgenii.vinogradov@kuleuven.be) received the Dipl. Engineer degree from Saint-Petersburg Electrotechnical University (Russia) and a Ph.D. degree from UCLouvain (Belgium). His doctoral research was focused on multidimensional stochastic channel modeling. He is currently an associate researcher with the Electrical Engineering Department, KU Leuven. His research interests include such 6G technologies as Non-Terrestrial and Cell-Free Networks.

\end{document}